\documentclass[namedreferences,hyperref,optionalrh]{spr-sola}

\usepackage{graphicx}        
\usepackage{amsmath}
\usepackage[percent]{overpic}
\usepackage{url}
\usepackage{color}           

\usepackage{hyperref} 

\newcommand{\aap}{{\it Astron. Astrophys.}}

\newcommand{\apj}{{\it Astrophys. J.}}
\newcommand{\apjl}{{\it Astrophys. J. Lett.}}

\newcommand{\grl}{{\it Geophys. Res. Lett.}}

\newcommand{\jgr}{{\it J. Geophys. Res.}}

\newcommand{\solphys}{{\it Sol. Phys.}}
 
\newcommand{\ssr}{{\it Space Sci. Rev.}} 
\chardef\us=`\_

\DeclareUnicodeCharacter{2500}{\textemdash}
\begin{document}
\hypersetup{
  colorlinks=true,
  citecolor=blue,
  linkcolor=blue,
  urlcolor=blue
}
\begin{frontmatter}
\title{Visualizing the Magnetic Structure in Interplanetary Coronal Mass Ejections with ATHARV}

\author[addressref={aff1},email={vivekmenon@iisc.ac.in}]{\inits{V.M.}\fnm{Vivek}~\snm{Menon}\orcid{0009-0007-8208-3960}}
\author[addressref={aff2,aff3},corref,email={sheoran.j03@gmail.com}]{\inits{J.}\fnm{Jyoti}~\snm{Sheoran}\orcid{0000-0003-1713-119X}}
\author[addressref=aff4,corref,email=vpant@iitd.ac.in]{\inits{V.}\fnm{Vaibhav}~\snm{Pant}\orcid{0000-0002-6954-2276}}
\author[addressref={aff5,aff6}]{\inits{D.}\fnm{Dipankar}~\snm{Banerjee}\orcid{0000-0003-4653-6823}}

\address[id=aff1]{Bachelor of Science (Research) Programme, Indian Institute of Science (IISc), Bengaluru, 560012, India}
\address[id=aff2]{Aryabhatta Research Institute of Observational Sciences (ARIES), Nainital - 263002, Uttarakhand, India}
\address[id=aff3]{Department of Applied Physics, Mahatma Jyotiba Phule Rohilkhand University, Bareilly - 243006, Uttar Pradesh, India}
\address[id=aff4]{Indian Institute of Technology Delhi, Hauz Khas, Delhi-110016, India}
\address[id=aff5]{Indian Institute of Space Science and Technology (IIST), Valiamala, Thiruvananthapuram - 695547, Kerala, India}
\address[id=aff6]{Center of Excellence in Space Sciences India (CESSI), IISER Kolkata, Mohanpur - 741246, West Bengal, India}

\runningauthor{Menon et al.}
\runningtitle{ATHARV: A Tool for Spatial Reconstruction of ICME Magnetic Structure }

\begin{abstract}

Interplanetary coronal mass ejections (ICMEs) are major drivers of space weather, and their geoeffectiveness is strongly governed by the structure and orientation of their internal magnetic field. However, in-situ observations provide only one-dimensional sampling of  magnetic-field properties along the spacecraft trajectory, limiting direct inference of the three-dimensional ICME magnetic structure. We introduce the \textit{Analysis Tool for Heliospheric Arrangement of Remapped Vectors} (ATHARV), which remaps in-situ time-series data into spatial coordinates while accounting for ICME expansion and spacecraft motion. The framework assumes self-similar expansion with different expansion rates along three orthogonal directions, while more general cases use measured velocities as proxies for plasma motion. ATHARV also incorporates complementary diagnostics, including hodograms and magnetic-field orientation angles, to assess magnetic coherence and field rotation within ICMEs.  We demonstrate the application of ATHARV using multipoint in-situ observations of an ICME detected near 1~au by STEREO-A and Wind on 2023 April 23--24. The reconstructed sheath region exhibits variable and disordered magnetic fields, whereas the magnetic ejecta (ME) shows a coherent rotation consistent with a south--west--north flux-rope configuration with right-handed helicity at both spacecraft. However, differences in the magnetic-field magnitude profiles, rotation signatures, and inferred ME sizes between the two spacecraft suggest mesoscale inhomogeneity within the ICME magnetic configuration, possibly associated with a writhed or distorted flux-rope geometry. This event highlights the limitations of interpreting the magnetic configuration of ICMEs from single-point measurements alone and demonstrates the importance of multipoint observations for investigating their three-dimensional structure and evolution. Overall, ATHARV provides a consistent framework for interpreting in-situ ICME observations and investigating their spatial structure and evolution. ATHARV is available online for the heliophysics community.

\end{abstract}
\end{frontmatter}

\section{Introduction}\label{S-Introduction}

Coronal Mass Ejections (CMEs; \citealt{2012_webb}) are large-scale eruptions of magnetized plasma expelled from the solar corona into the heliosphere, where they are observed in situ as Interplanetary Coronal Mass Ejections (ICMEs). These transients are among the primary drivers of heliospheric variability and space-weather disturbances, and their interaction with planetary magnetospheres can lead to severe geomagnetic activity \citep{2012_Mostl,2017a_Kilpua}. The geoeffectiveness of an ICME depends critically on the strength, orientation, and spatial extent of its internal magnetic field, particularly the presence of a sustained southward component that couples efficiently to the terrestrial magnetosphere \citep{1988_Tsurutani,2008_Echer}. A detailed understanding of the three-dimensional magnetic structure and its evolution during interplanetary propagation is, therefore, essential for both fundamental heliophysics and space-weather forecasting.

In situ observations commonly reveal an ICME as a composite structure consisting of a forward shock, a turbulent sheath, and a magnetic ejecta (ME). Fast ICMEs drive a forward shock that compresses and heats the upstream solar wind, forming a sheath region characterized by enhanced density, temperature, plasma beta, and strong magnetic field fluctuations \citep{2017b_Kilpua}. The ME represents the core magnetic structure of the ICME and is frequently interpreted as a magnetic flux rope (MFR),
although some events exhibit multiple flux ropes or lack clear flux-rope signatures altogether \citep{2003_Cane, 2006_Jian, 2010_Richardson, 2019_nieves, 2025_Al-haddad}. The ME region generally exhibits enhanced and coherent magnetic fields, reduced plasma beta, and depressed density and temperature \citep{2006_Zurbuchen}. A subset of MEs, referred to as magnetic clouds (MCs), display a smooth rotation of the magnetic field vector, often through angles approaching 180°, consistent with spacecraft traversals through an MFR \citep{1981_Burlaga, 1986_Marubashi, 1988_Burlaga}. The MFR has a coherent magnetic field structure in which helical magnetic field lines wrap around a central axis \citep{1981_Burlaga, Goldstein1983, 1986_Marubashi}. However, such idealized signatures are observed in only a fraction of ICMEs ($\sim$30\%), with occurrence rates varying from $\sim$15\% at solar maximum to $\sim$60\% at solar minimum \citep{Gosling1990,2003_Cane}. Despite this diversity, flux-rope ICMEs are often associated with the most intense geomagnetic storms, owing to their ability to sustain strong southward magnetic fields over extended intervals \citep{2011_Wu}.

As ICMEs propagate through interplanetary space, they continuously interact with the structured solar wind, leading to changes in their size, speed, and geometry. Following the initial expansion phase close to the Sun, revealed by coronagraph observations in which the CME leading edge propagates faster than its core \citep{2006_Tripathi}, the later radial evolution of ICMEs is primarily governed by pressure balance with the ambient solar wind \citep{2009_Demoulin}. The radial expansion of the ICME flux rope or ejecta (ME) is primarily driven by the decrease in the ambient solar-wind pressure with heliocentric distance and can often be approximated as self-similar under typical solar-wind conditions \citep{2009_Demoulin,2010_Gulisano}. Other factors, such as internal over-pressure, the presence of a shock, and the radial distribution and degree of twist within the flux rope, generally play a secondary role in governing the expansion \citep{2009_Demoulin}. Observationally, in-situ measurements commonly show a decrease in the bulk plasma velocity from the front to the rear of the ME, providing direct evidence of its expansion in the solar wind (\citealt{Leitner2007}, and references therein). The magnitude of the expansion of ICME ejecta is quantified using the velocity difference between the leading and trailing edges of the ejecta. However, the expansion of the ME is not always clearly discernible in in-situ plasma velocity measurements. Small or slow magnetic ejecta, particularly those propagating at speeds comparable to the ambient solar wind, often exhibit weak or poorly defined expansion signatures \citep{Tsurutani2004}. In addition, interactions with trailing high-speed streams or CME–CME interactions can substantially modify the expansion, leading to compression of the rear portion of the ejecta, flattened velocity profiles, or even locally negative expansion \citep{Burlaga2003,2013_Lugaz}. Both multispacecraft observations and numerical MHD simulations demonstrate that such interactions can suppress the nominal expansion or even produce locally negative expansion signatures in the ME \citep{Xiong2007, 2012_Ruffenach, 2013_Lugaz}, highlighting the crucial role of the ambient solar-wind environment in controlling the radial evolution of ICMEs.

The magnetic configuration of an ICME flux rope is often described using classification schemes based on chirality and axis orientation \citep{1998_Bothmer,1998_Mulligan}. In ideal magnetohydrodynamics, magnetic helicity is conserved, implying that the chirality of a flux rope is expected to remain unchanged during its propagation through interplanetary space \citep{1984_Berger,2007_Demoulin}. In contrast, the global orientation of the flux rope is not a topological invariant and may evolve as the ICME interacts with its surrounding environment. While a purely self-similar expansion would preserve the flux-rope orientation \citep{2009_Demoulin,2010_Gulisano}, observations and data-constrained modeling indicate that significant rotation and deflection can occur due to interactions with the structured solar wind, CME–CME interactions, and large-scale magnetic forces \citep{2014_Isavnin,2013_Lugaz,2021_Winslow}. Multi-spacecraft investigations further show that flux-rope orientations inferred at different heliocentric distances or longitudes within the same event can differ substantially, reflecting either the physical evolution of the structure or spatial distortions sampled at different locations \citep{2019_Good,2020_Davies,2025_Maunder}. In addition, magnetic reconnection between the ICME and the ambient solar wind can erode portions of the flux rope, modifying its magnetic configuration and further complicating the interpretation of its global structure \citep{2007_Dasso,2015_Ruffenach}.

Despite extensive in situ measurements, ICME vector quantities are traditionally presented as separate time series of individual components, which inherently mask their three-dimensional magnetic structure and temporal evolution. Furthermore, the  measured velocity profile during an ICME traversal depend not only on the intrinsic expansion of the flux rope but also on the relative motion between the spacecraft and the structure \citep{2003_Lynch, 2008_Demoulin}, implying that spacecraft motion can influence the inferred expansion speeds and radial sizes. This effect becomes increasingly important for missions with significant orbital velocities, such as Parker Solar Probe, which can traverse a non-negligible distance during an ICME encounter. Motivated by these considerations, we introduce theAnalysis Tool for Heliospheric Arrangement of Remapped Vectors (\textit{ATHARV}), a visualization framework that remaps in situ vector measurements of magnetic field into three-dimensional quiver representations in space and time by explicitly accounting for expansion and spacecraft position while preserving the original time-series information.  By enabling direct visualization of ICME magnetic-field geometry, \textit{ATHARV} provides new insight into ICME size, expansion, orientation, handedness, and internal substructure, and offers a complementary approach for interpreting single- and multi-spacecraft observations. The article is organized as follows: In Section~\ref{sec:CME_viz}, we describe the \textit{ATHARV} visualization tool and outline its underlying assumptions and methodology. In Section~\ref{sec:application}, we present the results from the application of \textit{ATHARV} to in-situ observations of an ICME.  In Section~\ref{summary}, we discuss the main results, present the conclusions, and outline the scope for future improvements to ATHARV.

\begin{figure}[h!!]
    \centering
    \includegraphics[width=0.9\linewidth]{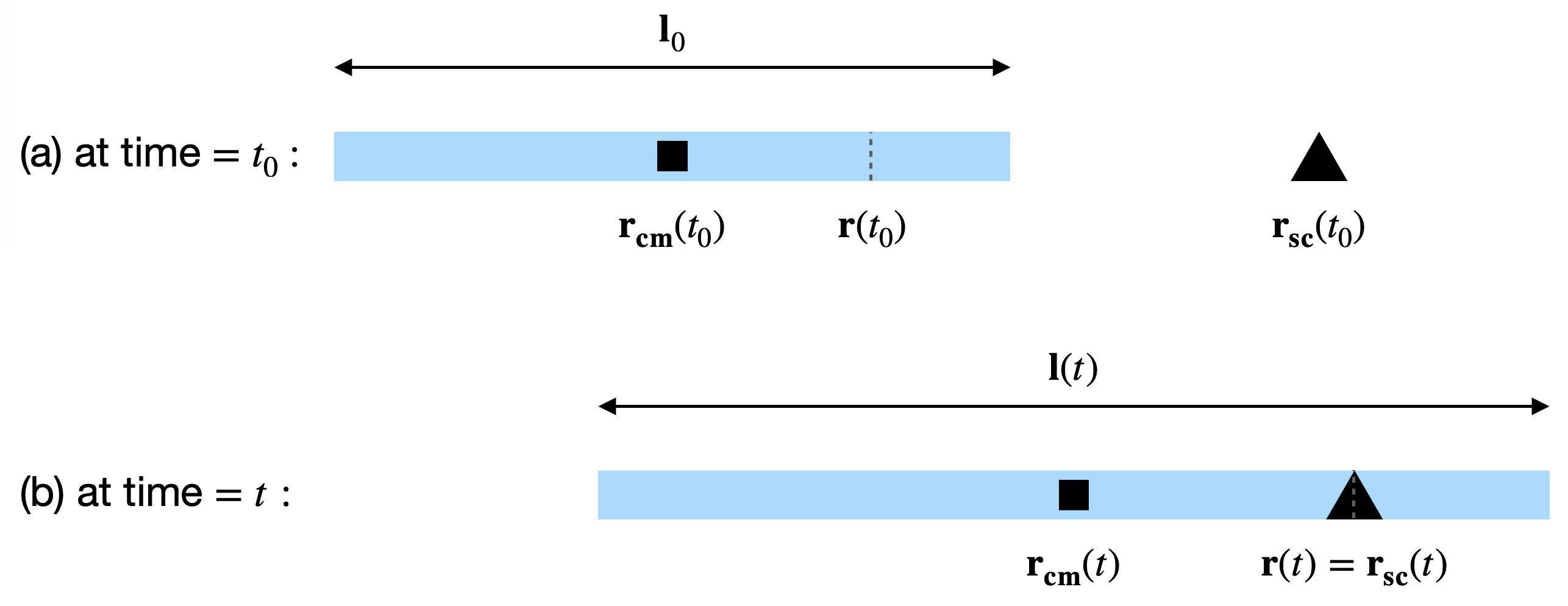}
    \caption{Schematic illustration of a self-similarly expanding one-dimensional structure (blue rectangle) at the reference time $t_0$ (top panel) and at a later time $t$ (bottom panel), propagating toward increasing coordinate values (to the right). The structure evolves from size $\mathbf{l_0}(t_0)$ centered at $\mathbf{r}_{\mathrm{cm}}(t_0)$ to size $\mathbf{l}(t)$ centered at $\mathbf{r}_{\mathrm{cm}}(t)$. The spacecraft (black triangle) performing in-situ measurements is located at $\mathbf{r}_{\mathrm{sc}}(t_0)$ at time $t_0$. A plasma parcel initially located at $\mathbf{r}(t_0)$ evolves to $\mathbf{r}(t)$ and is observed when it reaches the spacecraft position $\mathbf{r}_{\mathrm{sc}}(t)$. The remapping procedure aims to recover the original parcel position at the reference time $t_0$.}
    \label{fig:diagram}
\end{figure}

\section{ATHARV: CME Visualization Tool} \label{sec:CME_viz}

In-situ spacecraft observations of an ICME consist of time-series measurements of plasma and magnetic-field parameters recorded along the spacecraft trajectory through the structure. As the spacecraft traverses the ICME, the structure continues to expand while propagating outward, causing plasma parcels within the ejecta to convect at different speeds. Therefore, the observed temporal profiles reflect a combination of the intrinsic spatial structure of the ICME, its expansion during the observation interval, and the motion of the spacecraft through the evolving structure. The ATHARV visualization tool reconstructs the spatial magnetic structure along the spacecraft path by remapping the measured time series into spatial coordinates while accounting for ICME expansion and spacecraft motion. Within this framework, the ICME is treated as an ensemble of plasma parcels in which the magnetic field is assumed to be frozen into the plasma and does not evolve appreciably over the duration of the spacecraft passage. Figure~\ref{fig:diagram} illustrates the reconstruction geometry. At a reference time $t_0$, the ICME is represented as a one-dimensional structure of size $\mathbf{l_0}(t_0)$, centered at position $\mathbf{r}_{\mathrm{cm}}(t_0)$ in a non-rotating inertial frame, while the spacecraft is located at $\mathbf{r}_{\mathrm{sc}}(t_0)$. Each plasma parcel occupies a position $\mathbf{r}$ within the interval $\mathbf{r}_{\mathrm{cm}} \pm \mathbf{l_0}/2$. The structure propagates in the positive direction (to the right), evolves to a size $\mathbf{l}(t)$ at time $t$, and its centre moves to $\mathbf{r}_{\mathrm{cm}}(t)$, while the spacecraft samples successive parcels as they convect past it. A parcel initially located at $\mathbf{r}(t_0)$ is observed at time $t$ at position $\mathbf{r}(t)$ when it reaches the spacecraft position $\mathbf{r}_{\mathrm{sc}}(t)$. The objective is to determine the parcel position $\mathbf{r}(t_0)$ at the reference time $t_0$. The remapping procedure is described below.

\subsection{Case A: Uniform Positive Expansion}\label{sec:exp_correct}

ICME ejecta often exhibit velocity profiles in which the leading edge propagates faster than the trailing edge, indicating expansion during the spacecraft traversal. In such cases, we assume a self-similar (uniform) expansion, allowing for different expansion rates along three orthogonal directions \citep{2008_Demoulin}. The expansion in each direction is assumed to be governed by the corresponding measured velocity component. Under the assumption of uniform expansion, plasma parcels move away from the structure center at a constant rate. Accordingly, the structure expands with velocity $\mathbf{v}_{\mathrm{exp}}$, while its center propagates with a constant velocity $\mathbf{v}_{\mathrm{cm}}$.

Let the leading and trailing edges of the ICME be observed at times $t_{\mathrm{LE}}$ and $t_{\mathrm{TE}}$, with corresponding velocities $\mathbf{v}_{\mathrm{LE}}$ and $\mathbf{v}_{\mathrm{TE}}$. The center and expansion velocity 
are obtained from a linear fit to the in-situ velocity profile between these times, such that

\begin{equation}
\mathbf{v}_{\mathrm{LE}} = \mathbf{v}_{\mathrm{cm}} + \mathbf{v}_{\mathrm{exp}}, \qquad
\mathbf{v}_{\mathrm{TE}} = \mathbf{v}_{\mathrm{cm}} - \mathbf{v}_{\mathrm{exp}}.
\label{eq:boundary_velocities}
\end{equation}

Since the velocity varies linearly across the structure, the velocity at position $\mathbf{r}$ is

\begin{equation}
\mathbf{v}(\mathbf{r}) =
\mathbf{v}_{\mathrm{cm}}
+ \frac{2\mathbf{v}_{\mathrm{exp}}}{\mathbf{l}_0}
(\mathbf{r} - \mathbf{r}_{\mathrm{cm}}).
\label{eq:linear_velocity_profile}
\end{equation}

If a parcel located at $\mathbf{r}$ at time $t_0$ is observed at spacecraft position $\mathbf{r}_{\mathrm{sc}}(t)$ at time $t$, the parcel displacement satisfies

\begin{equation}
\mathbf{r}_{\mathrm{sc}}(t) - \mathbf{r}
=
\left(
\mathbf{v}_{\mathrm{cm}}
+ \frac{2\mathbf{v}_{\mathrm{exp}}}{\mathbf{l}_0}
(\mathbf{r} - \mathbf{r}_{\mathrm{cm}})
\right)(t - t_0).
\label{eq:parcel_displacement}
\end{equation}

Solving for the parcel position at the reference time $t_0$ yields

\begin{equation}
\mathbf{r} =
\frac{
\mathbf{r}_{\mathrm{sc}}(t)
- \mathbf{v}_{\mathrm{cm}}(t-t_0)
+ \dfrac{2\mathbf{v}_{\mathrm{exp}}\mathbf{r}_{\mathrm{cm}}}{\mathbf{l}_0}(t-t_0)
}{
1 + \dfrac{2\mathbf{v}_{\mathrm{exp}}}{\mathbf{l}_0}(t-t_0)
}.
\label{eq:expansion_mapping}
\end{equation}

This relation remaps each time-series measurement to its spatial position within the ICME at $t_0$. The unknown parameters $\mathbf{r}_{\mathrm{cm}}$ and $\mathbf{l}_0$ are determined from boundary conditions defined by the observed arrival times of the leading and trailing edges and the spacecraft positions.
The above procedure is applied component-wise in the RTN coordinate system, with propagation and expansion velocities evaluated independently along the $R$, $T$, and $N$ directions using the respective velocity components ($V_R$, $V_T$, $V_N$).

\subsection{Case B: Non-uniform or Negative Expansion}\label{sec:simple_viz}

If the in-situ velocity profile of the ICME structure indicates negative expansion ($v_{\mathrm{LE}} < v_{\mathrm{TE}}$) or does not exhibit a linearly declining trend from the leading to the trailing edge, 
as may occur in ICME sheaths \citep{2021_Salman} or dynamically perturbed ejecta \citep{2008_Demoulin, 2010_Gulisano}, the uniform expansion model is not applicable. Such velocity profiles can arise from local compression, overtaking high-speed streams, CME–CME interactions, shocks, or pressure-gradient forces  \citep[for eg; See][]{burlaga2003merged, 2007_Xiong, 2020_Regnault, 2021_Salman}.

In these cases, the measured in-situ velocities are assumed to approximate the Lagrangian velocities of plasma parcels convecting past the spacecraft. If a parcel located at $\mathbf{r}$ at the reference time $t_0$ is observed at the spacecraft position $\mathbf{r}_{\mathrm{sc}}(t)$ at time $t$, then

\begin{equation}
\mathbf{r} =
\mathbf{r}_{\mathrm{sc}}(t) -
\int_{t_0}^{t} \mathbf{v}(t')\, dt'.
\label{eq:lagrangian_mapping}
\end{equation}

This approach traces parcel trajectories backward in time to estimate their positions at $t_0$. It is consistent with the assumption of a frozen-in magnetic structure and is most appropriate over short-duration intervals. However, when applied to expanding regions, it may lead to an overestimation of spatial scales.

\subsection{Application to ICME Sheath and Magnetic Ejecta}

To obtain the remapped positions of plasma parcels within the ICME, the reference time $t_0$ is defined as the arrival of the sheath leading edge. The sheath trailing edge reaches the spacecraft at time $t_1$, marking the onset of the ME, while the trailing edge of the ME is observed at time $t_2$.
For plasma parcels sampled within the sheath interval ($t_0 \le t \le t_1$), remapped positions are obtained using Equation~(\ref{eq:lagrangian_mapping}). The same approach is adopted for MEs that are locally contracting or exhibit strongly distorted or non-linear velocity profiles, in which case Equation~(\ref{eq:lagrangian_mapping}) is applied over the full interval ($t_0 \le t \le t_2$).

If the ME exhibits a clear expansion signature, remapped positions are instead computed using Equation~(\ref{eq:expansion_mapping}). To ensure continuity at the sheath–ME boundary, the remapped position of the sheath trailing edge, which coincides with the leading edge of the ME, is given by

\begin{equation}
\mathbf{r}_{\mathrm{LE,~ME}} =
\mathbf{r}_{\mathrm{TE,~sheath}} =
\mathbf{r}_{\mathrm{sc}}(t_0)
-
\int_{t_0}^{t_1} \mathbf{v}(t')\, dt' = \mathbf{r}_{\mathrm{sc}}(t_0) - \mathbf{l}_{\mathrm{0,~sheath}}.
\label{eq:boundary_position}
\end{equation}

Assuming the trailing edge of the ME propagates at velocity $\mathbf{v}_{\mathrm{cm}}-\mathbf{v}_{\mathrm{exp}}$, it must traverse the combined thickness of the sheath and ejecta together with the spacecraft displacement during the encounter, such that

\begin{equation}
\mathbf{l}_{0,~\mathrm{ME}} + \mathbf{l}_{0,~\mathrm{sheath}}
+ \mathbf{r}_{\mathrm{sc}}(t_2) - \mathbf{r}_{\mathrm{sc}}(t_0)
=
(\mathbf{v}_{\mathrm{cm}}-\mathbf{v}_{exp})(t_2 - t_0).
\label{eq:me_thickness_relation}
\end{equation}

The initial ME size is therefore

\begin{equation}
\mathbf{l}_{0,\mathrm{ME}} =
(\mathbf{v}_{\mathrm{cm}}-\mathbf{v}_{exp})(t_2 - t_0)
- \mathbf{l}_{0,\mathrm{sheath}}
- \mathbf{r}_{\mathrm{sc}}(t_2)
+ \mathbf{r}_{\mathrm{sc}}(t_0).
\label{eq:me_size}
\end{equation}

The ME center position at $t_0$ is

\begin{equation}
\mathbf{r}_{cm,\mathrm{ME}} =
\mathbf{r}_{\mathrm{sc}}(t_0)
-
l_{0,~\mathrm{sheath}}
-
\frac{\mathbf{l}_{0,~\mathrm{ME}}}{2}.
\label{eq:me_center}
\end{equation}

With these quantities determined, Equations~(\ref{eq:expansion_mapping}) and (\ref{eq:lagrangian_mapping}) can be used to map the position of each plasma parcel within the ICME at the reference time $t=t_0$.

\section{Application to in-situ data} \label{sec:application} 

In this section, we demonstrate the application of the ATHARV tool\footnote{\url{https://vivekmenon42.github.io/ATHARV}} (see Appendix~\ref{sec:viz_app} for detailed instructions on operating the web interface) to in-situ observations of an ICME detected by STEREO-A and Wind on 2023 April 23–24 and illustrate its capability to infer its spatial extent and magnetic configuration. At the time of observation, the spacecraft were located at heliocentric distances of 0.96 ~au (STEREO-A) and 1.00~au (Wind), with heliographic coordinates (longitude, latitude) of ($-10.2^\circ$, $-5.8^\circ$) and ($0.0^\circ$, $-4.9^\circ$), respectively.  This spacecraft configuration, characterized by small radial and longitudinal separations, enables mesoscale sampling of the ICME structure, which is typically achievable when multiple spacecraft are separated by $\sim$0.005–0.05 au radially and $\sim$1$^\circ$–12$^\circ$ longitudinally \citep{2018_Lugaz}. 

Proton plasma parameters for STEREO-A are obtained from PLASTIC \citep{Galvin2008, Luhmann2008}, with magnetic-field data from IMPACT/MAG \citep{Acuna2008}. For Wind, we use proton plasma measurements from SWE \citep{Ogilvie1995} and magnetic-field measurements from MFI \citep{Lepping1995}. All data are expressed in the radial–tangential–normal (RTN) coordinate system, where $\mathbf{\hat{R}}$ is directed radially outward from the Sun, $\mathbf{\hat{T}} = \mathbf{\hat{\Omega}{\odot}} \times \mathbf{\hat{R}}$, with $\mathbf{\hat{\Omega}{\odot}}$ denoting the solar rotation axis, and $\mathbf{\hat{N}}$ completes the right-handed system. 

\begin{figure}[t]
\centering

\begin{overpic}[width=0.49\textwidth]{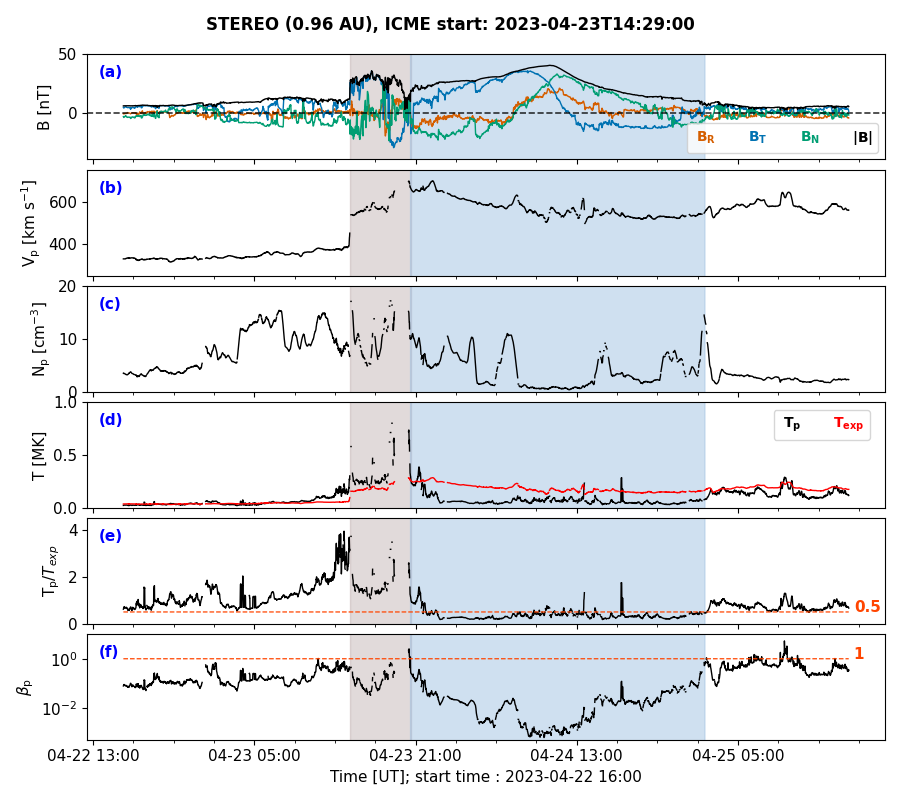}
\put(50,-5){\textbf{(i)}}
\end{overpic}
\hfill
\begin{overpic}[width=0.49\textwidth]{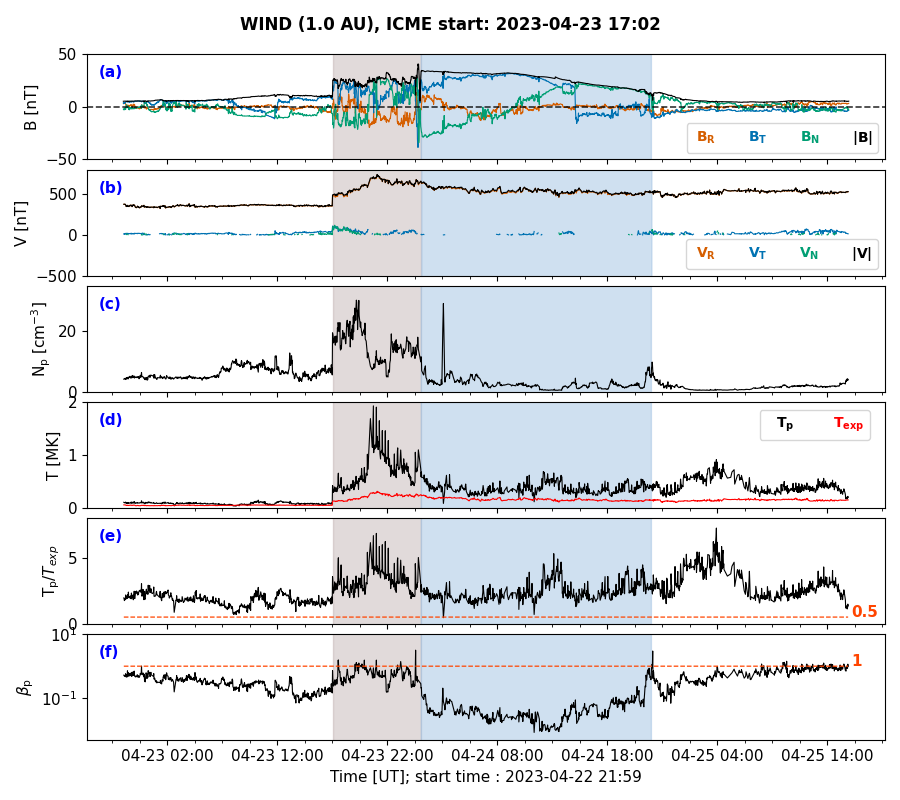}
\put(50,-5){\textbf{(ii)}}
\end{overpic}

\vspace{0.4cm}
\caption{Time series of magnetic-field and plasma parameters observed by (i) STEREO-A and (ii) Wind during the ICME event of 23–24 April 2023. The panels show: (a) magnetic-field components ($B_R$, $B_T$, $B_N$) and total magnitude ($|B|$); (b) proton velocity components ($V_R$, $V_T$, $V_N$) and bulk speed ($V_p$); (c) proton number density ($N_p$); (d) proton temperature ($T_p$, blue) and the expected temperature ($T_{\mathrm{exp}}$, red) derived from the empirical $V_p$–$T_p$ relationship; (e) the ratio $T_p/T_{\mathrm{exp}}$ (the dashed red line indicates $T_p/T_{\mathrm{exp}} = 0.5$); and (f) proton plasma beta ($\beta_p$, with the dashed red line marking $\beta_p = 1$). The ICME sheath and ME intervals are shaded light gray and light blue, respectively. For STEREO-A, proton velocity component data are not available.}
\label{fig:wind_sta_20230423}
\end{figure}

Figure~\ref{fig:wind_sta_20230423} presents the time-series measurements of magnetic-field and plasma parameters for the ICME observed by STEREO-A and Wind on 2023 April 23–24. Each panel shows the magnetic-field components ($B_R$, $B_T$, $B_N$) and their magnitude ($|B|$), together with the proton velocity components ($V_R$, $V_T$, $V_N$) and magnitude ($V_p$), density ($N_p$), temperature ($T_p$), the temperature ratio $T_p/T_{\mathrm{exp}}$, and plasma beta ($\beta_p$), where $T_{\mathrm{exp}}$ is the expected temperature derived from the empirical $V_p$–$T_p$ relationship \citep{1986_Lopez,Liu2005}. The panel layout is identical for both spacecraft, with the sheath region highlighted in light gray and the ME in light blue. The observations at both spacecraft exhibit a typical ICME structure, consisting of a turbulent sheath followed by a well-defined ME interval. The bulk-speed profile across the ME shows a gradual decline from the leading to trailing edge, indicating ongoing expansion of the structure during propagation. A linear fit to the ME bulk-speed profile yields expansion speeds of 69.14~km~s$^{-1}$ at STEREO-A and 40.43~km~s$^{-1}$ at Wind. Since the ME exhibits a clear expansion signature at both locations, the \textit{expansion-corrected remapping} technique is applied to the ATHARV reconstructions within the ME interval.

\begin{figure}[t]
\centering

\begin{overpic}[width=0.9\textwidth]{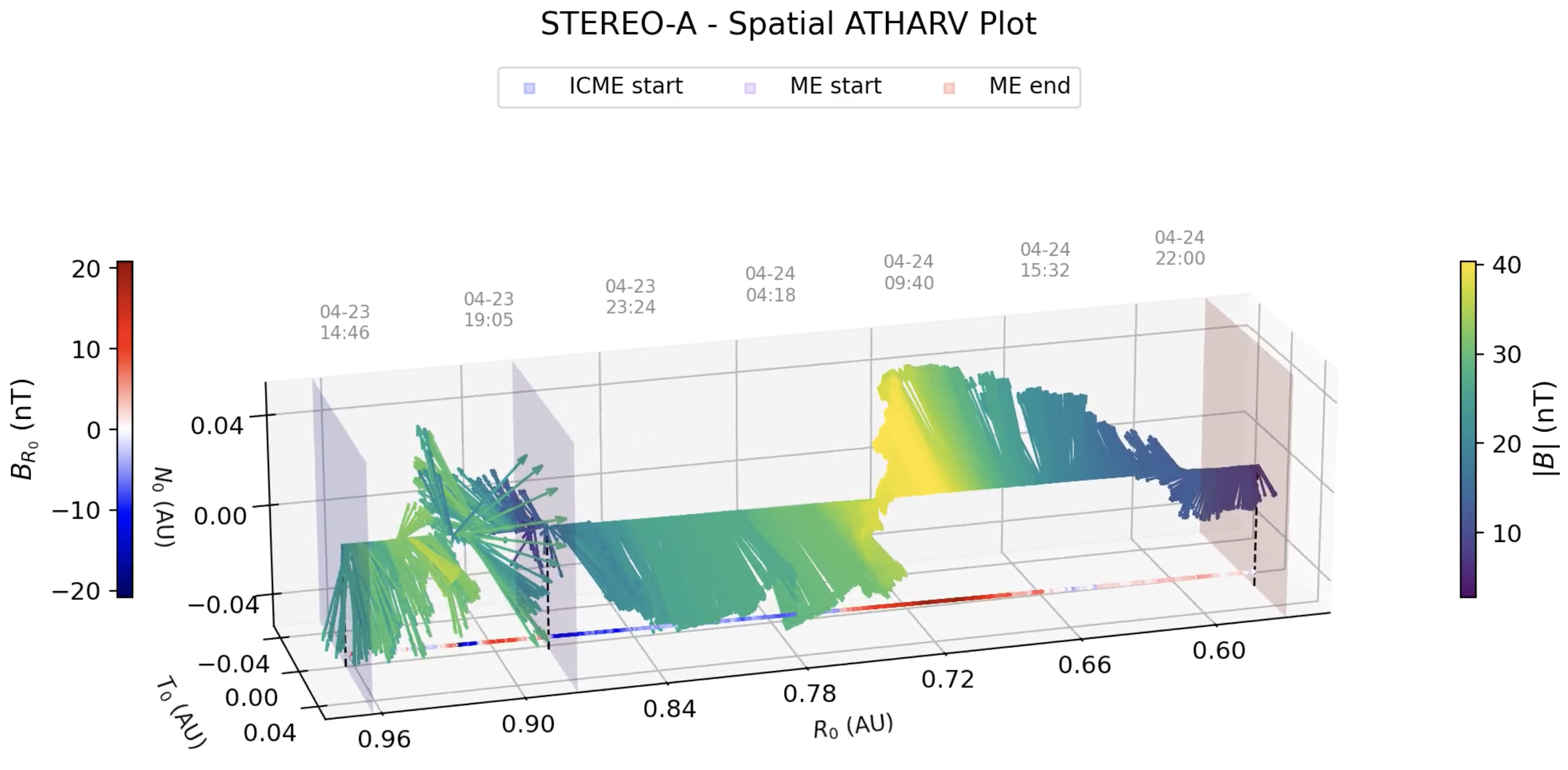}
\put(-2,47){\textbf{(i)}}
\end{overpic}

\vspace{0.4cm}

\begin{overpic}[width=0.9\textwidth]{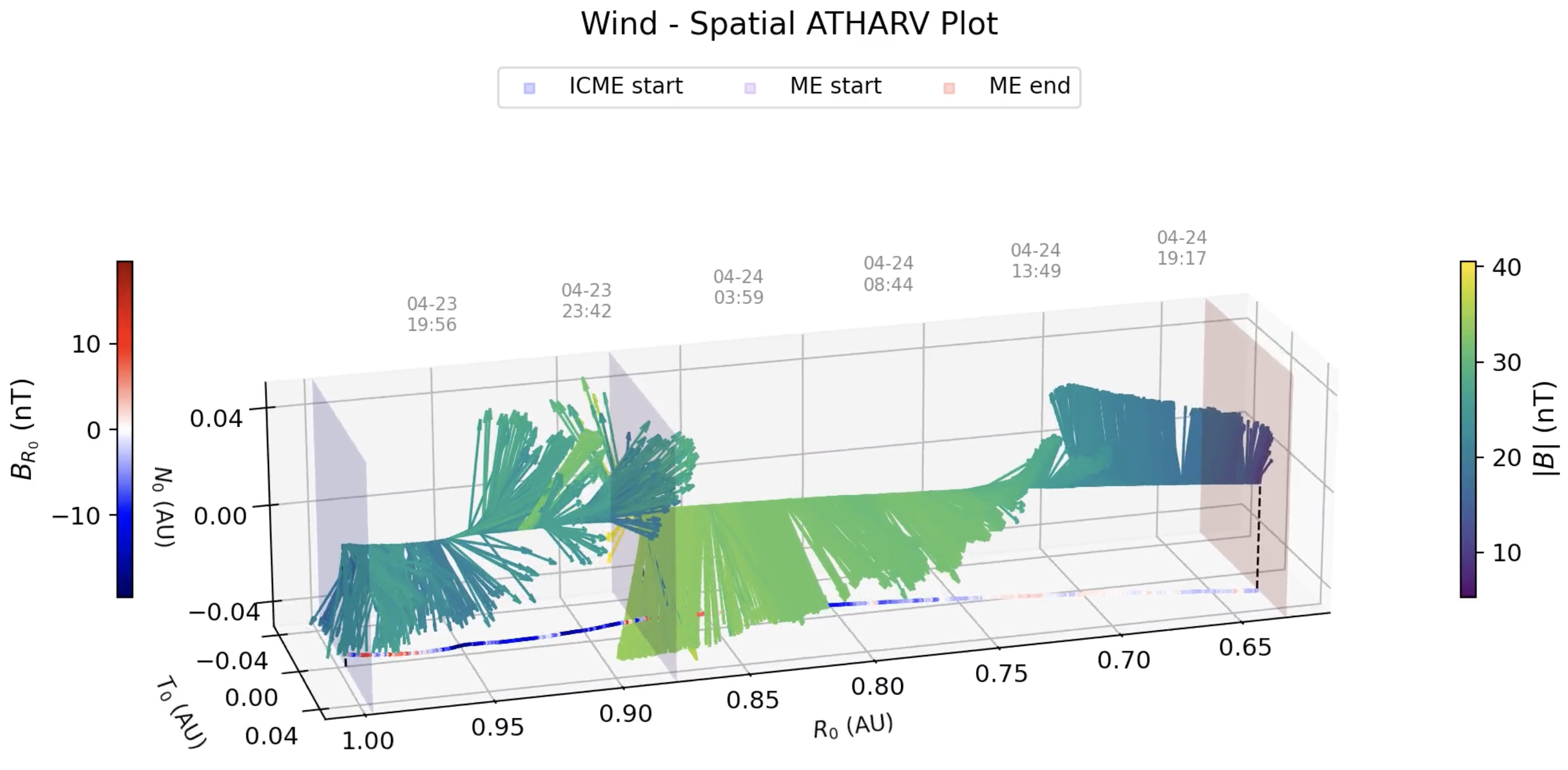}
\put(-2,47){\textbf{(ii)}}
\end{overpic}

\caption{Three-dimensional reconstruction of the ICME magnetic field structure along the spacecraft trajectories during the ICME event of 2023 April 23–24 for (i) STEREO-A and (ii) Wind, generated using the ATHARV tool. Arrows indicate the magnetic field direction, with length and color representing $|B|$ in the $R_{0}T_{0}N_{0}$ frame at the initial ICME encounter time $t_{0}$. The arrow tails trace the remapped spacecraft trajectory in the $R_{0}T_{0}N_{0}$ frame, illustrating the path sampled through the ICME. A parallel red--white--blue strip shows the magnitude and sign of the $B_{R_{0}}$ component along the same trajectory. The ICME propagates radially outward (to the left); the blue, indigo, and red planes mark the ICME start, ME start, and ME end, respectively. Animations of this figure, showing snapshots from multiple viewing angles, are available as Electronic Supplementary Material at \url{https://doi.org/10.6084/m9.figshare.32274126}.}
\label{fig:atharv_sta_wind}
\end{figure}

Figure~\ref{fig:atharv_sta_wind} presents the ATHARV reconstructions of the ICME on 2023 April 23–24 at the STEREO-A and Wind locations, providing a three-dimensional visualization of the magnetic-field vectors within the ICME. The reconstruction is performed in a non-rotating heliocentric coordinate system, $R_0T_0N_0$, defined such that its axes are aligned with the spacecraft RTN frame at the initial ICME encounter time $t_0$ (e.g., $t_0 = 2023$-04-23 14:29 UT for STEREO-A and $t_0 = 2023$-04-23 17:02 UT for Wind). The magnetic-field vectors are represented by arrows whose orientation indicates the local field direction, while their length and color encode the field magnitude ($|B|$). The arrow tails are anchored along the remapped spacecraft trajectory in the $R_0T_0N_0$ frame, illustrating the path sampled through the ICME. In addition, a parallel red–white–blue strip denotes the magnitude and sign of the $B_{R_0}$ component along the trajectory. The ICME propagates radially outward (towards the left in Figure~\ref{fig:atharv_sta_wind}), with the blue, indigo, and red planes marking the start of the sheath, the onset of the ME, and the end of the ME, respectively. The corresponding remapped spatial coordinates of these boundaries are used to estimate the radial extents of the sheath and ME. The sheath size is found to be $\sim$0.085~au at STEREO-A and $\sim$0.12~au at Wind, while the ME size is $\sim$0.28~au at STEREO-A and $\sim$0.25~au at Wind, indicating modest differences in ICME sheath and ME size at both spacecraft locations. For comparison, conventional estimates based on the average speed and event duration yield sheath radial sizes of 0.083~au at STEREO-A and 0.12~au at Wind, and ME radial sizes of 0.40~au at STEREO-A and 0.27~au at Wind. While the sheath radial sizes are comparable to the ATHARV-based estimates, the conventionally estimated ME radial sizes are systematically larger. This indicates that conventional approaches can overestimate the true radial size of expanding MEs. The discrepancy is substantially larger at STEREO-A, where the ME exhibits a stronger expansion signature, suggesting that higher expansion speeds lead to longer observed durations and consequently to a significant overestimation of the true radial size of the ME.

The reconstructed magnetic-field vectors clearly distinguish the sheath and ME regions. The sheath is characterized by highly variable and disordered magnetic-field vectors, consistent with enhanced turbulence in the compressed upstream region. In contrast, the ME exhibits a smooth and coherent rotation of the magnetic field along with gradual variations in magnitude, indicative of an organized flux-rope configuration. Notably, throughout the ME, the magnetic-field vector rotates from southwest to northeast through the westward direction in RTN coordinates, corresponding to a south--west--north (SWN) type flux rope following the classification of \citet{1998_Bothmer}. The sense of rotation indicates a right-handed (positive helicity) flux rope with a moderate inclination relative to the Sun--spacecraft plane. Here, “north” refers to the direction parallel to the $N$ axis, while “west” corresponds to the positive $T$ direction. Although the overall magnetic structure is consistent between the two spacecraft, the magnetic-field magnitude exhibits a more symmetric profile at STEREO-A, increasing toward the center of the ejecta and decreasing toward the trailing edge, whereas at Wind it shows a more gradual decline across the ME.

\begin{figure}[htbp]
\centering

\begin{minipage}{0.48\textwidth}
    \centering
    \begin{overpic}[width=\linewidth]{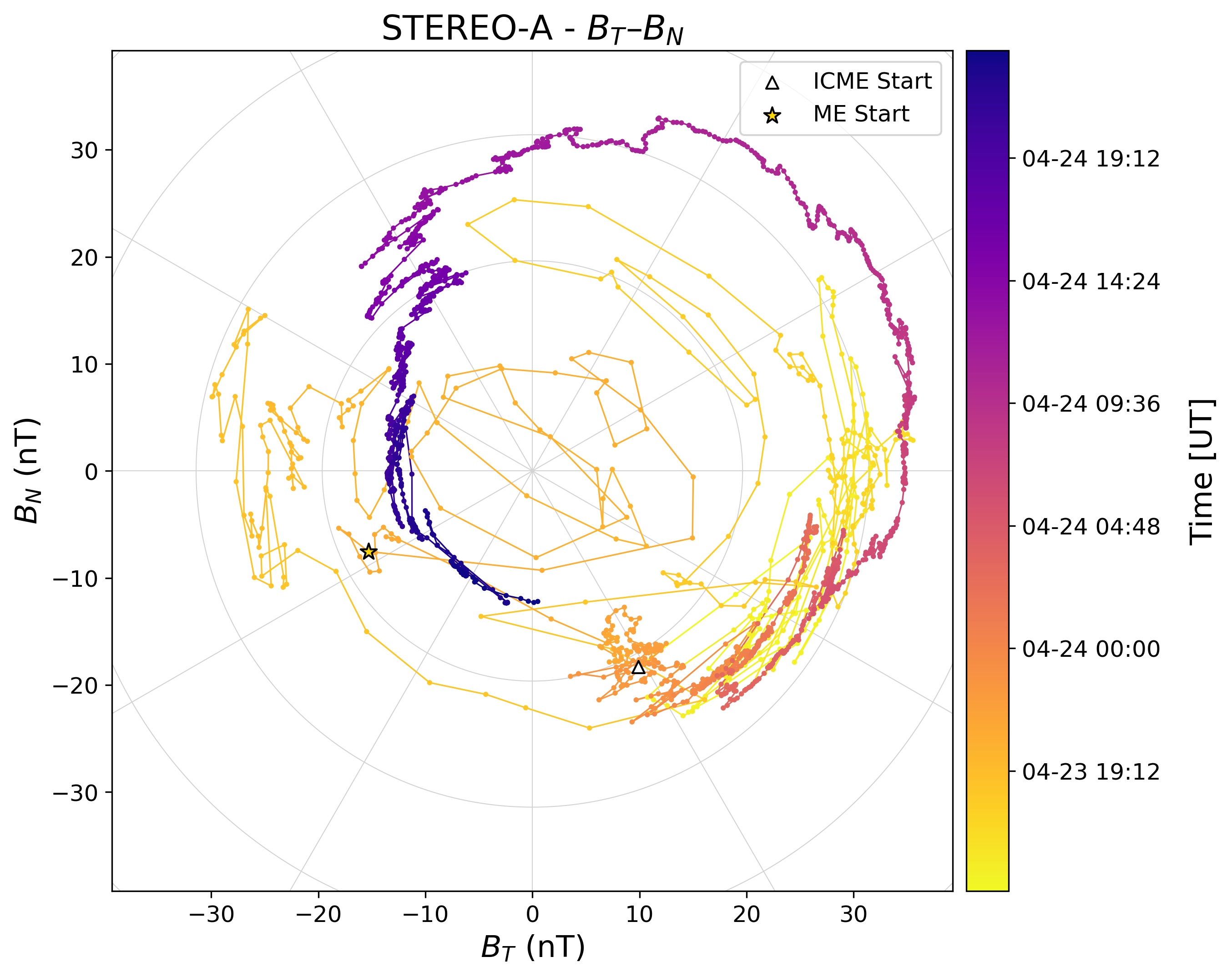}
        \put(-3,79){\textbf{(i)}}
    \end{overpic}
\end{minipage}
\hspace{0.02\textwidth}
\begin{minipage}{0.48\textwidth}
    \centering
    \begin{overpic}[width=\linewidth]{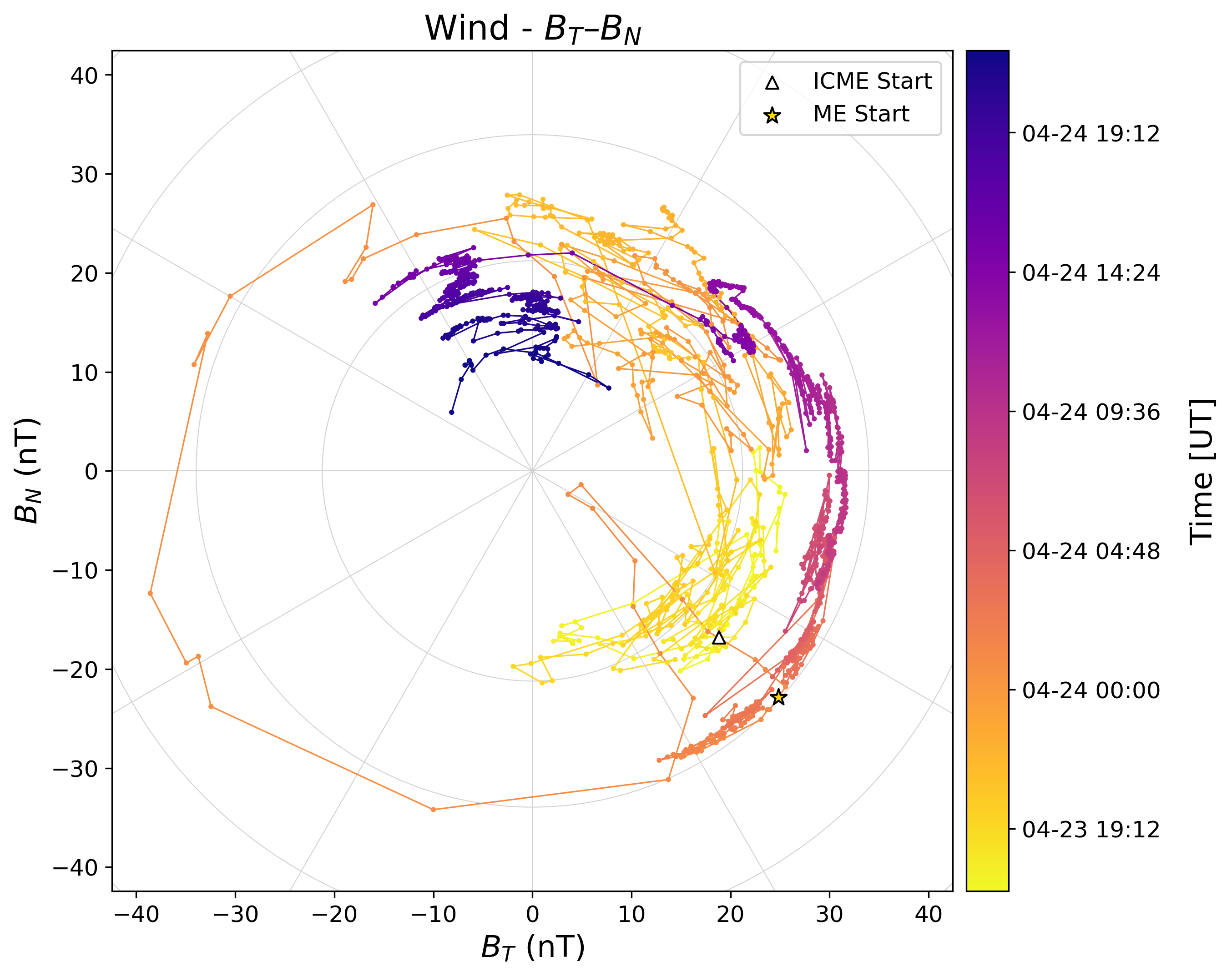}
        \put(-3,79){\textbf{(ii)}}
    \end{overpic}
\end{minipage}

\vspace{0.3cm}

\begin{minipage}{0.48\textwidth}
    \centering
    \begin{overpic}[width=\linewidth]{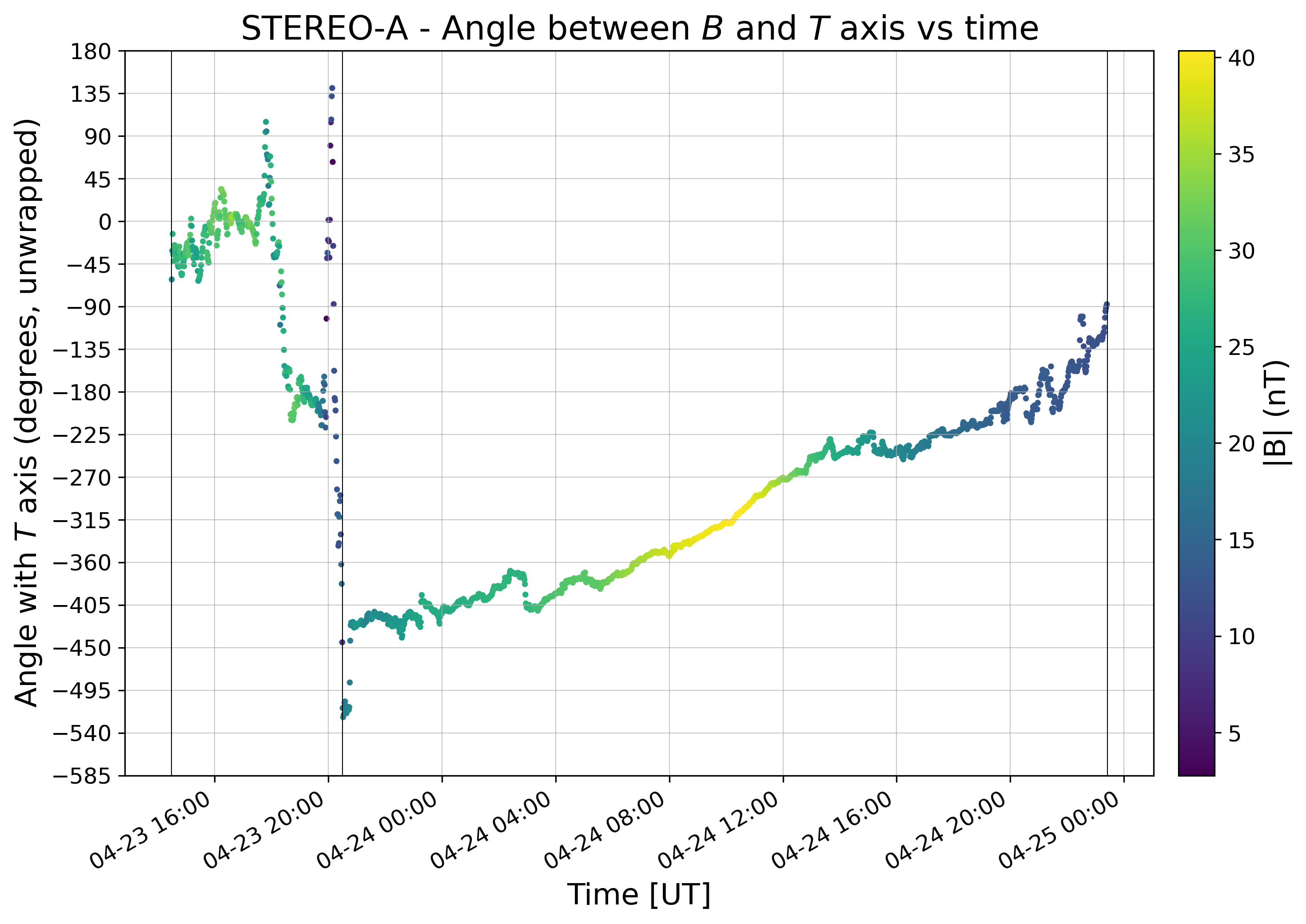}
        \put(-2,72){\textbf{(iii)}}
    \end{overpic}
\end{minipage}
\hspace{0.02\textwidth}
\begin{minipage}{0.48\textwidth}
    \centering
    \begin{overpic}[width=\linewidth]{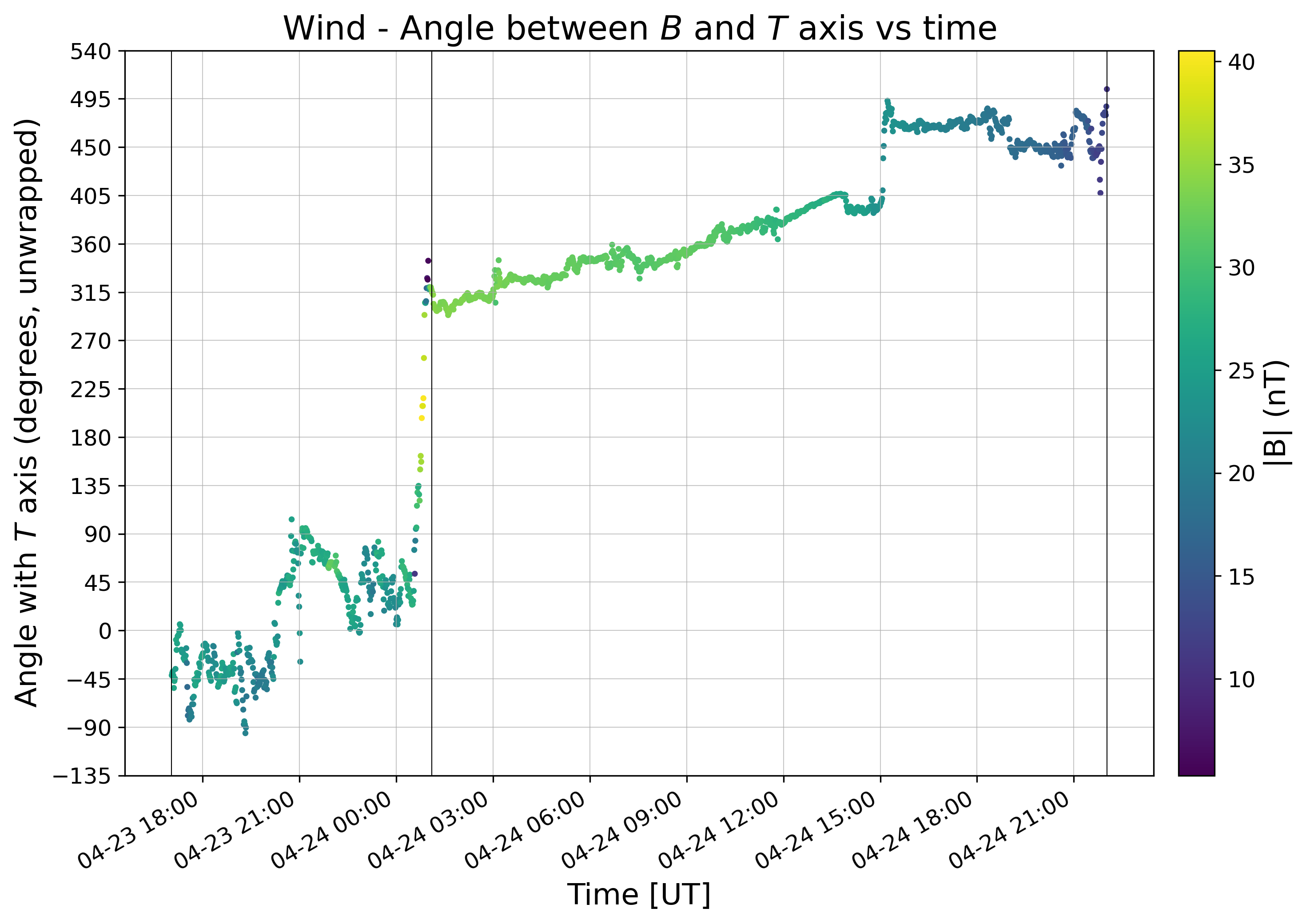}
        \put(-2,72){\textbf{(iv)}}
    \end{overpic}
\end{minipage}

\caption{Top panels: Magnetic-field hodograms ($B_T$–$B_N$) for the ICME event of 2023 April 23–24 observed at (i) STEREO-A and (ii) Wind. Data points and connecting lines are color-coded by observation time. The white triangle marks the ICME onset, and the gold star indicates the onset of the ME. Bottom panels: Temporal evolution of the magnetic-field vector angle with respect to the $T$-axis at (iii) STEREO-A and (iv) Wind. Data points are color-coded by magnetic-field magnitude, and vertical lines indicate the ICME and ME boundaries.}
\label{fig:sensu_angle}

\end{figure}

Figure~\ref{fig:sensu_angle} presents the $B_T$--$B_N$ hodograms and the magnetic-field vector angle with respect to the $T$-axis, which together provide complementary diagnostics of the magnetic-field rotation. The top panel shows the $B_T$--$B_N$ hodograms of the ICME at the STEREO-A and Wind locations, with data points and connecting lines color-coded by observation time. The white triangle marks the ICME onset, and the gold star indicates the onset of the ME. In the sheath region, the magnetic field exhibits irregular and highly variable rotations, consistent with turbulent fluctuations in the compressed plasma. In contrast, the ME region displays a coherent rotation, with the hodograms forming a semicircular to near-circular arc in the $B_T$--$B_N$ plane, providing clear evidence of a flux-rope structure. The ME hodogram at STEREO-A exhibits a more extended arc compared to Wind, suggesting mesoscale inhomogeneity or internal complexity within the magnetic structure of the ejecta across its angular extent.
The bottom panel shows the temporal evolution of the magnetic-field vector angle with respect to the $T$-axis, with data points color-coded by magnetic-field magnitude and vertical lines marking the ICME and ME boundaries. In the sheath, the angle is dominated by random fluctuations, reflecting the disordered magnetic field. Within the ME, however, the angle increases smoothly with time, indicating a systematic rotation of the magnetic-field vector.  At Wind, the total angle change is approximately $180^\circ$, consistent with a single traversal of a flux rope. In contrast, at STEREO-A, the angle exceeds $180^\circ$ (reaching $\sim330^\circ$), indicating a double rotation within the ME, i.e., in addition to the primary $\sim180^\circ$ rotation, a secondary rotation is present.

\section{Discussion and Conclusions} \label{summary} 

ICMEs are key drivers of heliospheric variability and space-weather disturbances, with their geoeffectiveness largely governed by the strength, orientation, and spatial extent of their internal magnetic field. However, in-situ observations are inherently limited to time-series measurements along the spacecraft trajectory, providing only one-dimensional sampling of the ICME structure and thereby restricting direct inference of its three-dimensional configuration and evolution. Furthermore, conventional approaches for estimating ICME size based on duration and average velocity do not fully account for the effects of ICME expansion and spacecraft motion on the measured velocity profile, leading to potential ambiguities in the inferred structure. These limitations become particularly significant for fast-moving missions such as the Parker Solar Probe.

In this work, we introduce \textit{ATHARV}, a visualization and analysis framework that reconstructs the spatial structure of ICMEs by remapping in-situ time-series measurements into spatial coordinates while accounting for ICME expansion and spacecraft motion. For expanding structures, the method incorporates a self-similar framework that allows for anisotropic expansion along three orthogonal directions, consistent with previous studies \citep{2008_Demoulin}. The framework is adaptable to a range of observational conditions. In cases where expansion is not well defined, the measured in-situ velocities are treated as proxies for the Lagrangian motion of plasma parcels. In addition to spatial reconstruction, ATHARV provides complementary diagnostics such as hodograms and magnetic-field orientation angles relative to the T-axis, facilitating a more comprehensive assessment of magnetic coherence and rotation within the ICME. The ATHARV tool thus provides a more physically consistent representation of ICME structure and enables improved characterization of their size, magnetic configuration, and internal variability.

We present an application of the ATHARV tool to multipoint in-situ observations of an ICME detected near 1~au by STEREO-A and Wind on 2023 April 23–24, demonstrating its capability to infer the spatial extent and magnetic configuration of the ejecta. The sheath and ME sizes are estimated to be \(\sim0.085~\mathrm{au}\) and \(\sim0.28~\mathrm{au}\) at STEREO-A, and \(\sim0.12~\mathrm{au}\) and \(\sim0.25~\mathrm{au}\) at Wind, respectively. At both spacecraft, the sheath is characterized by highly variable and disordered magnetic fields, whereas the ME exhibits a smooth and coherent rotation consistent with a south--west--north (SWN) type flux rope with right-handed helicity. However, notable differences are observed in the magnetic-field magnitude profiles and rotation signatures. At Wind, the magnetic-field magnitude gradually decreases across the ME interval and the total magnetic-field rotation is close to \(180^\circ\), broadly consistent with a classical magnetic-cloud configuration represented by a cylindrical, force-free flux rope with a circular cross-section and a nearly straight invariant axis \citep{1982_Burlaga,1983_Goldstein,1988_Burlaga,1990_Lepping}. However, the observed gradual decline in magnetic-field magnitude is not fully reproduced by such idealized force-free models, although it is consistent with the asymmetric magnetic-field profiles commonly observed in magnetic clouds near 1~au. Previous studies have attributed these asymmetries in the magnetic-field magnitude profile within the ME to expansion and/or aging effects \citep{1993_Farrugia,1993_Osherovich}. More recent studies, however, suggest that the asymmetry of the magnetic-field profile shows little dependence on the ME expansion speed \citep{2018_Nieves} and may instead reflect intrinsic structural asymmetries within the flux rope. In contrast, the comparatively symmetric magnetic-field profile at STEREO-A suggests a crossing closer to the central region of the flux rope, where the magnetic-field strength is expected to be higher. The large rotation angle approaching \(\sim330^\circ\) at STEREO-A, nevertheless, cannot be readily explained by a simple force-free cylindrical flux rope with a straight invariant axis. One possible interpretation is that the ejecta possessed a writhed or distorted flux-rope geometry during heliospheric propagation, similar to the updated schematic representation proposed by \citet[see their Figure~10]{2025_Al-haddad}. In such a configuration, the flux-rope axis itself becomes curved or rotationally distorted, producing substantial variations in the local axis orientation along the spacecraft trajectory. Consequently, the observed magnetic-field rotation may include contributions from both the intrinsic helical twist of the magnetic field lines and the large-scale curvature of the flux-rope axis, allowing rotations substantially larger than \(180^\circ\) \citep{2022_Weiss}. The comparatively larger ME size and stronger expansion signature observed at STEREO-A further suggest that the spacecraft sampled a more distorted portion of the ejecta, whereas Wind intersected a relatively simpler section of the structure. Additional effects such as asymmetric expansion, interaction with the ambient solar wind, or localized magnetic erosion \citep{2007_Dasso,2012a_Ruffenach,2015_Ruffenach} may have further contributed to the observed mesoscale differences.

Overall, these results suggest that the ME has a non-coherent magnetic structure at mesoscales across its angular extent, such that local variations in flux-rope geometry can produce substantially different in-situ signatures even for spacecraft separated by only \(\sim10^\circ\) in longitude. In this context, the ejecta observed at STEREO-A resembles a ``complex ejecta'', whereas the Wind observations are more consistent with a ``simple ejecta'', following the classification proposed by \citet{2025_Al-haddad}. The STEREO-A observations exhibit multiple magnetic-field rotations (approaching \(\sim360^\circ\)), together with a comparatively larger radial size (approximately twice the typical MC size (\(\sim0.21~\mathrm{au}\)) at 1~au) and a stronger expansion signature, characteristics commonly associated with complex ejecta. Note that the larger ME size discussed here corresponds to the conventional duration-based estimate (\(\sim0.40~\mathrm{au}\) at STEREO-A), which exceeds both the ATHARV-based estimate and the typical MC size. This event therefore highlights the limitations of interpreting the global magnetic configuration of ICMEs from single-point measurements alone and emphasizes the importance of multipoint observations for characterizing the three-dimensional structure and evolution of heliospheric flux ropes.

These findings demonstrate the capability of ATHARV to investigate the spatial structure and magnetic complexity of ICMEs using multipoint in-situ observations. The ATHARV framework therefore provides a practical tool for improving the interpretation of in-situ ICME observations and supports a more reliable characterization of ICME size and magnetic structure, which is important for advancing our understanding of ICME evolution and their impact on the near-Earth environment. The tool is available online for the heliophysics community. Future developments will focus on integrating ATHARV with existing ICME catalogs, such as HELIO4CAST \citep{2017_Mostl, 2020_Mostl, 2022_Mostl}, enabling users to directly select events and generate standardized spatial reconstructions across multiple spacecraft. We also plan to incorporate data from recent missions, including Aditya-L1 \citep{2017_Seetha, 2023_Tripathi}, whose in-situ instruments—the Magnetometer \citep[MAG;][]{2025_Vipin} and the Aditya Solar Wind Particle Experiment \citep[ASPEX;][]{2025_Kumar, 2025_Goyal}—will provide continuous measurements at L1 and complement existing observations of ICMEs near Earth. In addition, ATHARV will be extended to support user-provided datasets for spacecraft whose observations are not currently accessible through standard archives such as CDAWeb. These developments will facilitate more systematic investigations of ICME evolution across heliocentric distances and support the broader application of ATHARV in future studies of ICME structure and heliocentric evolution.

\begin{acks}

The authors would like to express their sincere gratitude to Dr. Satabdwa Majumdar (Austrian Space Weather Office, GeoSphere Austria) and Dr. Ritesh Patel (Southwest Research Institute, Boulder) for their valuable insights and constructive suggestions. The authors also thank Dhruv Dua (Undergraduate Student at the Indian Institute of Science Education and Research, Bhopal, India) for constructive input during the initial phase of developing the website. V.M. acknowledges the support of the NIUS programme of HBCSE–TIFR, funded by the Department of Atomic Energy, Government of India (Project No. RTI4001). This work has made use of the ICME LineupCAT hosted at \url{https://helioforecast.space/lineups}, developed by Christian Möstl and collaborators as part of the HELIO4CAST project. The authors also acknowledge the use of NASA’s Goddard Space Flight Center (GSFC) Space Physics Data Facility’s Coordinated Data Analysis Web (CDAWeb) for access to the data utilized in this study.

\end{acks}

\begin{authorcontribution}
V.M. conceptualized and developed the ATHARV framework, including the methodology, software implementation, and web interface, and carried out the formal analysis and investigation. J.S. contributed to the development and application of ATHARV, identified the ICME event for demonstration, prepared the figures, and wrote the original manuscript draft. V.M. prepared the ATHARV visualization figures and animations. J.S., V.P., and D.B. contributed to the interpretation and discussion of the results. V.P. assisted with manuscript review and editing, and D.B. acquired funding support for V.M. All authors contributed to the scientific discussion and approved the final manuscript.
\end{authorcontribution}

\begin{fundinginformation}
V.M. acknowledges support from the NIUS programme of HBCSE--TIFR, funded by the Department of Atomic Energy, Government of India (Project No.~RTI4001). J.S. is supported by the Department of Science and Technology, Government of India, for research at ARIES.
\end{fundinginformation}

\begin{dataavailability}
All data used in our study is publicly accessible via NASA CDAWeb (\href{https://cdaweb.gsfc.nasa.gov}{https://cdaweb.gsfc.nasa.gov})
\end{dataavailability}

\begin{materialsavailability}
Electronic supplementary material, including ATHARV animations of the ICME reconstruction from multiple viewing angles, is available at \url{https://doi.org/10.6084/m9.figshare.32274126}
\end{materialsavailability}

\begin{codeavailability}
The ATHARV source code is publicly available at \url{https://github.com/vivekmenon42/ATHARV.git}. The interactive ATHARV web interface is available at \url{https://vivekmenon42.github.io/ATHARV}. 
\end{codeavailability}

\begin{ethics}
\begin{conflict}
The authors declare that there are no conflicts of interest.
\end{conflict}
\end{ethics}

\appendix
\renewcommand{\thefigure}{A\arabic{figure}}
\setcounter{figure}{0}  
\renewcommand{\thetable}{A\arabic{table}}
\setcounter{table}{0}

\begin{figure}[h!!]
    \centering
    \begin{overpic}[width=\linewidth]{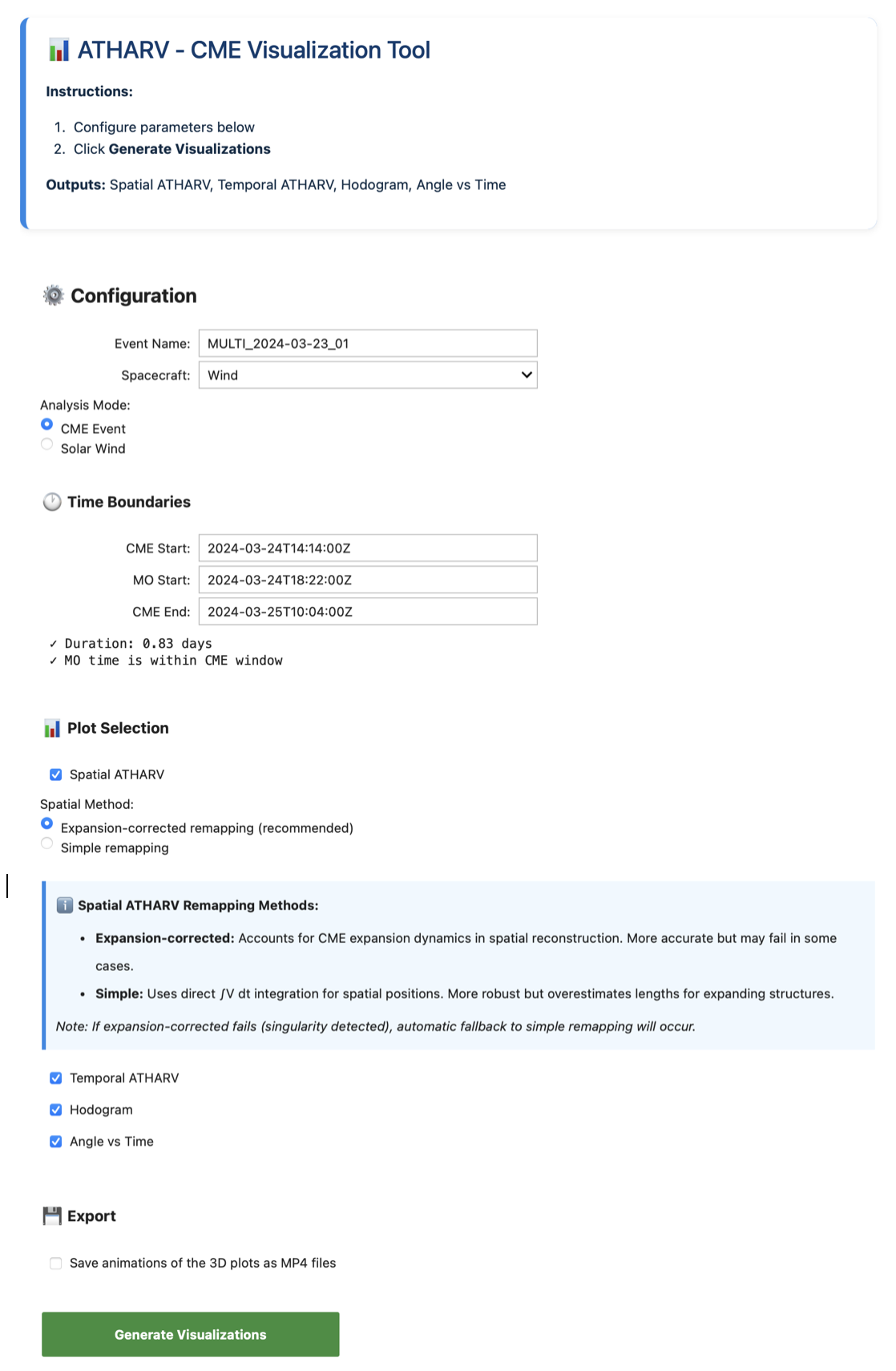}
        \put(-1,75){\textbf{1}}
        \put(-1,72.5){\textbf{2}}
        \put(-1,70){\textbf{3}}
        \put(-1,62.8){\textbf{4}}
        \put(-1,46.5){\textbf{5}}
        \put(-1,10.8){\textbf{6}}
    \end{overpic}
    
    \caption{Basic features of ATHARV: (1–2) define event name and select spacecraft; (3) select analysis mode; (4) define time boundaries— for \textit{CME Event}, specify ICME/sheath start time, MO start time, and ICME/MO end time, and for \textit{Solar Wind}, specify the start and end times of the interval of interest; (5) select visualization options (\textit{Spatial ATHARV}, \textit{Temporal ATHARV}, \textit{Hodogram}, and \textit{Angle vs Time}), with choice of remapping method for \textit{Spatial ATHARV}; (6) export animations as MP4 files.}
    
    \label{fig:atharv_web}
\end{figure}

In this Appendix, we provide detailed instructions for operating the ATHARV web interface\footnote{\url{https://vivekmenon42.github.io/ATHARV}}. Figure~\ref{fig:atharv_web} presents a screenshot of the ATHARV web application, which integrates all functionalities of the tool into an interactive visualization environment.

\subsection{Operating the ATHARV Web Interface}\label{sec:viz_app}
After launching the tool through the interactive interface, the user first selects the ``Spacecraft" and then the ``Analysis Mode". Two modes are available: \textit{CME Event} and \textit{Solar Wind}.
For the \textit{CME Event} mode,
the user specifies three time boundaries: \textit{CME Start}, \textit{MO Start}, and \textit{CME End}. The \textit{CME Start} corresponds to the time when the ICME shock or sheath is encountered by the spacecraft, while \textit{MO Start} marks the beginning of the magnetic obstacle/ejecta (MO/ME) interval. The \textit{CME End} denotes the end of the MO region.

Under ``Plot Selection", four visualization options are provided to examine the ICME magnetic-field structure. The ``Spatial ATHARV" plot reconstructs the magnetic-field vectors in remapped spatial coordinates using the method described in Section~\ref{sec:CME_viz}. Two reconstruction approaches are available. The \textit{expansion-corrected remapping} (Section~\ref{sec:exp_correct}) accounts for the dynamical expansion of the CME during the spacecraft traversal, whereas the \textit{simple remapping} (Section~\ref{sec:simple_viz}) assumes  that the  velocity measured at the spacecraft approximates the Lagrangian velocity of plasma parcels within the CME to derive the remapped spatial positions. The ``Temporal ATHARV" plot displays the magnetic-field vectors as a function of time using the original spacecraft time-series measurements. In addition, the ``Hodogram ($B_T$--$B_N$)" option produces $B_T$--$B_N$ hodograms, which serve as a complementary diagnostic to assess the coherence and rotation of the magnetic field within the ICME, particularly in flux-rope ejecta \citep{1998_Bothmer, 2018_Nieves}. The final option, ``Angle vs Time", provides the temporal evolution of the magnetic-field angle with respect to the $T$-axis in the $T$--$N$ plane. A change of $360^\circ$ corresponds to one full rotation of the magnetic field vector, while larger cumulative angles indicate successive rotations.

To facilitate visualization of the three-dimensional magnetic-field structure, ATHARV also generates animations of the reconstructed 3D vector fields from multiple viewing angles. These animations help reveal the spatial structure and rotation of the magnetic field and can be exported as MP4 files. After selecting ``Generate Visualization", the ATHARV website retrieves the required in-situ spacecraft data from the CDAWeb database and produces the requested plots. The interactive interface further allows users to hover over vectors within the ATHARV plots to view information about the remapped spatial coordinates and the corresponding detection time at the spacecraft. The same visualizations---\textit{Spatial ATHARV}, \textit{Temporal ATHARV}, \textit{Hodogram}, and \textit{Angle vs Time}---are also available for selected solar-wind intervals using the \textit{Solar Wind} analysis mode. In this case, the ``Spatial ATHARV" reconstruction uses the \textit{simple remapping} technique, assuming that the solar-wind velocity measured at the spacecraft approximates the Lagrangian velocity of plasma parcels within the selected interval (Section~\ref{sec:simple_viz}).

\textbf{Note:} Linear interpolation is applied to gaps in the velocity measurements within the sheath or MO intervals. However, when large or frequent data gaps occur within the ICME interval, such interpolation may introduce significant uncertainties and should therefore be treated with caution. 
In addition, when velocity data are unavailable at the boundaries of the ICME interval, interpolation is not performed; instead, the earliest and latest available velocity measurements are extrapolated backward and forward, respectively, to enable remapping of the magnetic-field vectors at those boundaries.
Furthermore, in cases where only the bulk speed $V$ is available, and the individual components ($V_R$, $V_T$, $V_N$) are not measured, the spatial remapping is performed under the assumption that $V_R = V$ and $V_T = V_N = 0$. In cases where the velocity measurements are unavailable or contain frequent large gaps---particularly within the sheath region---or in ejecta that do not exhibit approximately linear expansion, the \textit{Spatial ATHARV} reconstruction may not be reliable. In such situations, the \textit{Temporal ATHARV} representation is preferred.



\end{document}